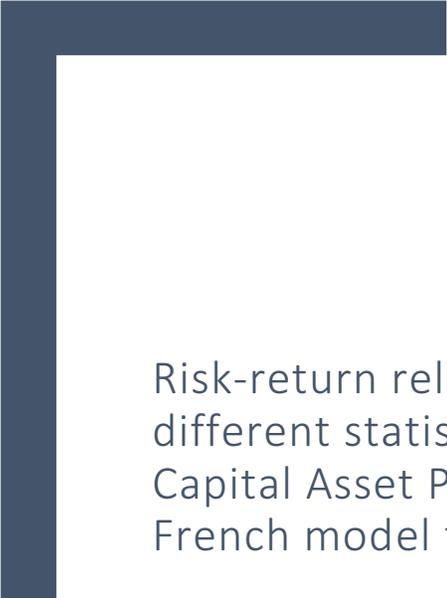

# Risk-return relationship: An empirical study of different statistical methods for estimating the Capital Asset Pricing Models (CAPM) and the Fama-French model for large cap stocks

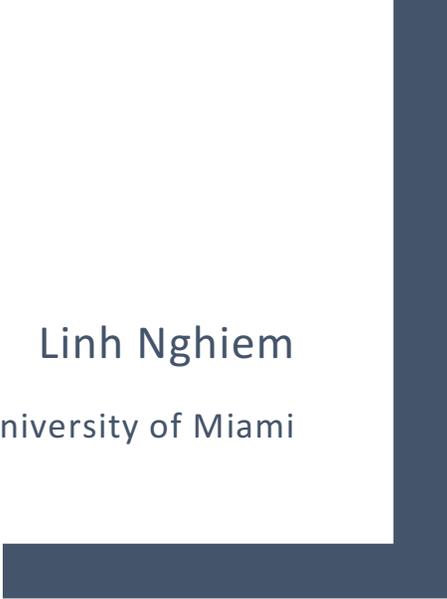

Linh Nghiem

University of Miami

# Contents

# LIST OF TABLES

# LIST OF FIGURES

# ABSTRACT

The Capital Asset Pricing Model (CAPM) is one of the original models in explaining risk-return relationship in the financial market. However, when applying the CAPM into reality, it demonstrates a lot of shortcomings. While improving the performance of the model, many studies, on one hand, have attempted to apply different statistical methods to estimate the model, on the other hand, have added more predictors to the model. First, the thesis focuses on reviewing the CAPM and comparing popular statistical methods used to estimate it, and then, the thesis compares predictive power of the CAPM and the Fama-French model, which is an important extension of the CAPM. Through an empirical study on the data set of large cap stocks, we have demonstrated that there is no statistical method that would recover the expected relationship between systematic risk (represented by beta) and return from the CAPM, and that the Fama-French model does not have a better predictive performance than the CAPM on individual stocks. Therefore, the thesis provides more evidence to support the incorrectness of the CAPM and the limitation of the Fama-French model in explaining risk-return relationship.

# INTRODUCTION

In the financial market, when one invests money into an asset, one hopes to get a bigger amount of money or an additional wealth sometime in the future. Investment, therefore, always includes a sacrifice of some resources today and an expectation of a greater benefit from them in the future. In other words, no matter what kind of assets one is investing, one expects return from it. The higher return, the better the investment. In addition, return of the most financial asset, including bonds, equities, derivatives, is directly derived from its price in the market. As prices vary from asset to asset, returns do; in the majority of case, because the price of an asset in the future is not exactly known at the time of investment, return is not either. Therefore, investors always attempt to find some shortcut to predict return of an asset to choose the asset into which they should invest.

Regardless of method of prediction, actual return always has the probability of being different, or even far different from expected, because prices are unpredictable. According to Kendall (1953) in his study of stock price, "the random changes from one term to the next are so large as to swamp any systematic effect which may be present"; in other words, stock prices followed no predictable pattern. As a matter of pure logic, given that price of a stock at some point *in the future* can be *precisely* predictable through current and past prices, everyone would sell or buy it, and the current price would jump immediately to the future price today, not in the future. The uncertainty of price, therefore, results in the uncertainty of return, so there always have some risks associated with each investment that indicates the volatility of its return. In fact, risk is one of the critical features that distinguish finance from other fields of economics.

Hence, a fundamental problem in both academic finance and in investment world is the risk-return relationship. If we are willing to take a high level of risk, does it guarantee a high return? In turn, is a high return always associated with a high level of risk? These questions are essential in choosing assets and building an investment portfolio, so many academic models were developed to examine risk-return relationship. The Capital Asset Pricing Model (CAPM), which describes the relationship by a simple linear regression, and the Fama-French three factor model (Fama-French model), a significant extension of the CAPM, are two most popular and recognized models. As always, any model is just a simplification of the real world through assumptions; nevertheless, the CAPM and Fama-French three factor model provide us with important understandings about risk and return in the financial market.

In the paper, we reviewed both the CAPM and the Fama-French model theoretically, including intuition, meaning and popular estimation methods. We then studied empirically how both models worked in a specific data set of big cap stocks, compared their predictive powers, and discussed about their feasibilities in the real world. Through reviewing them, we aimed to have better knowledge about risk and return, and thus provided valuable insights for investors to allocate wealth in their investment.

Structure of the paper as follows: Part II does some literature reviews on the CAPM and the Fama-French model. Part III reviews some basic measure of risk and return. Part IV and V review the Capital Asset Pricing Model (CAPM) and the Fama-French model theoretically, including their assumption, intuition and conclusion. Part VI reviews popular statistical methods in estimating the CAPM and the Fama-French model. Part VII is a comprehensive empirical study on a specific data set about the predictive power and feasibility of both models. Part VIII and IX are some discussions around the result of the empirical study. Part X is the conclusion of the thesis.

## LITERATURE REVIEW

The CAPM model was first developed by Sharpe, John Lintner, and Jan Mossin in 1967. It describes the relationship between expected excess return and systematic risk level (measured by beta coefficient) in a simple linear relationship:

$$E(R_i) - R_f = \beta \left( E(R_m) - R_f \right)$$

Recently, many papers have concluded that the CAPM model has not worked well in the real financial market both of the US and various countries: The beta coefficient is not a good predictor for return. (Alves 2013, Guzeldere & Sarioglu (2012), Soumaré et al (2013), etc.) In fact, many assumptions of the CAPM model do not hold in reality. One significant assumption is that the market portfolio could fully (or significantly) capture systematic risk; therefore, the beta of an asset, traditionally estimated by the ordinary least square OLS as the quotient of the covariance of stock's return with market portfolio's return and the variance of the market portfolio's return, is supposed to reflect this asset's systematic risk. Furthermore, this estimate is only efficient if many assumptions of the linear regression model are satisfied (independence, normality, etc.), which are also not usually the case in the financial market (Fama 1965). For example, return should be modeled by some heavily-tailed distribution, rather than normal distribution, because the probability of extreme return in the financial market are much higher than the probability estimated by the normal distribution (Fama 1965). For that reason, many papers also applied some other estimation methods in estimating the CAPM model, like the robust Bayesian method (Wong and Bian 2001), modified maximum likelihood (Bian, McAleer, & Wong 2013). Their empirical studies showed that the alternative methods were superior to the traditional OLS method in forecasting return from the CAPM model in some specific data sets.

Nevertheless, these papers did not reveal if the alternative methods used to estimate betas recovered the expected relationship from the CAPM model. In other words, with betas estimated with different methods, do stocks with high betas have higher returns than stocks with low betas? Investors in the financial market usually pay more attention to the relative return among assets (i.e, does stock A have a higher return than stock B?), rather than the absolute value of return (i.e, how much is the return of stock A?), it is important to consider if beta is still a good signal of return if estimated by different statistical methods.

## MEASURING RETURN AND RISK

Before we examine risk-return relationship, we need to define a consistent way to measure them. For the return, the most general formula for the total return of holding an asset/a portfolio over a period time, i.e. holding period return (HPR), is:

$$HPR = \frac{End\ of\ period\ value - Initial\ value + Income}{Initial\ value}$$

For an investor that invests in equity, the formula could be reduced to:

$$HPR_t = \frac{P_t - P_{t-1} + D_t}{P_{t-1}}$$

where $P_t$ and $P_{t-1}$ are equity prices at time t and t-1 respectively, while $D_t$ is dividend received from time t-1 to t.

The formula above gives the discretely-compounded return *($r_d$)* (discrete return), because here we have finite (1) compounding period (from t-1 to t). If, between the time t-1 to t, we have infinite number of compounding periods (each compounding period is infinitely short), we get continuously compounded return ($r_c$) (continuous return) given by taking the natural logarithms of the discrete return:

$$r_c = \log(1 + HPR_t) = \log(1 + r_d)$$

Note that these formulas above compute historical return, when we already know all prices and events in the past. However, when we buy the stock/portfolio at the time t-1, we only know $P_{t-1}$, but are unsure of how $P_t$ or $D_t$ would turn out; at time t-1, $P_t$ and $D_t$ are random variables, so is future return at time t (no matter return is continuously compounded or discretely compounded). We define expected future return as:

$$E(\tilde{r}_d) = \frac{E(P_t) - P_{t-1} + E(D_t)}{P_{t-1}}$$

$$E(\tilde{r}_c) = \log(1 + E(\tilde{r}_d))$$

Next, we define some measurement of risk. The total risk of an asset is usually implied through the volatility of historical returns, and often measured by the standard deviation of the past return:

$$\sigma = \sqrt{\frac{1}{m-1} \sum (r_{cm} - \bar{r}_c)^2}$$

While m is the number of the periods, $r_{cm}$ 's are historical discrete returns over period m, and $\bar{r}_c$ is the average of these historical returns. The formula for the risk of continuous return could be obtained by using continuous returns instead of discrete returns.

The general problem for investors as well as for academic finance is whether we could use information in the past to predict performance in the future. Related to the risk and return, two central questions are: (1) Can we predict future expected return based on historical information on risk and return? (2) Does there exist a strong correlation between risk and return (Is high return associated with high risk? Does a high risk guarantee a high return?) Many models that have been proposed and constantly improved to answer these questions, among which the Capital Asset Pricing Model (CAPM) and Fama-French model are the original and the most popular ones.

# THE CAPITAL ASSET PRICING MODEL (CAPM)
## Systematic and Idiosyncratic risk – Principle of Diversification

While standard deviation measures the total risk of an asset in the past, it says nothing about where the risk comes from, as well as how to control the level of risk. In fact, for each asset, its risk may come from a variety of sources, including economic conditions of the entire economy, conditions of the industry, government policy, and so forth. Nevertheless, in general, for each asset, risk factors fall into one of the two following categories:

- Systematic risk: As implied by its name, systematic risk is the risk that arises from the market structure and general economic conditions and more importantly, affects all agents in the market. Examples of systematic risk include business cycle, inflation, exchange rates. Nevertheless, sometimes whether risk is systematic depends on the broad mentioned context. For instance, US economy's business cycle is systematic for all US stocks, but it is not for international stocks that has nothing to do with the US. Because systematic risk has affected all agents, it is non-diversifiable, meaning that no matter what financial assets you hold, you are still exposed to the systematic risk.

- Idiosyncratic (or firm-specific) risk: Unlike the systematic risk, idiosyncratic risk is the risk that is only exposed by a specific firm/industry/agent. For example, weather may be a risk with companies in agriculture-related industry, like Archer Daniels Midland (ADM). While systematic risk could not be avoided, idiosyncratic risk could be reduced a lot through proper diversification, meaning that the action of holding a portfolio of assets of many risk-type assets rather than only one single risk-type asset.

The total risk of a financial asset, therefore, can be expressed as the sum of the systematic and idiosyncratic risk:

$$Total\ risk\ =\ Systematic\ Risk\ +\ Idiosyncratic\ risk$$

The graph below describes total risk and its components:

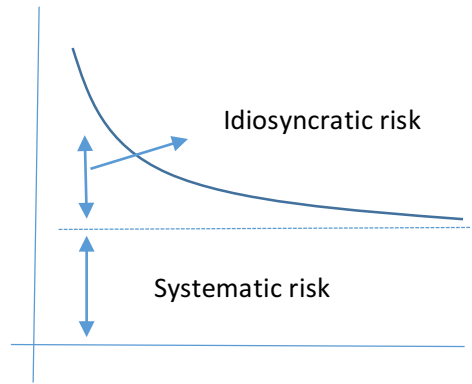

*Figure 1. Total Risk and its component.*

Because idiosyncratic risks can be avoided through proper diversification, it will not be compensated. A lot of papers and books have studies proper diversification, both in terms of security selection and asset allocation, and it is out of scope of the thesis. The thesis, therefore, only focused on the measurement of systematic risk, and the relationship between the systematic risk and expected return.

How do we measure systematic risk? This is a difficult question to answer. Asset pricing models attempt to use one or several risk factors, meaning that quantifiable indexes whose value tell us whether systematic risk is high or low. An intuition for risk factors come from the fact that assets that have higher payoff during "bad" times should be sold at a higher price, and thus have lower expected return. The risk factors would tell us when the "good" or "bad" times are;

for instance, the "market portfolio" considered a risk factor, which was used to develop the Capital Asset Pricing Model.

Because the systematic risk is unavoidable, it should be compensated, which results in *risk premium*. Asset risk premium is defined as the reward of bearing the risk; in other words, risk premium of an asset is the difference between the return of an asset and the risk-free rate.

## The Capital Asset Pricing Model

The Capital Asset Pricing Model (CAPM) is one of the most important models in modern financial economics; in fact, it was the benchmark of pricing securities and portfolio. The CAPM was first developed in 1967 by Sharpe, John Lintner, and Jan Mossin. The CAPM takes market portfolio as the only one risk factor into consideration. Market portfolio (usually denoted by M) is the portfolio that consists of every possible asset in the market with portfolio weights proportional to the market share (market cap) of each asset. The main idea of the CAPM is that, as the market portfolio is the most diversified portfolio, the expected risk premium of a financial asset should be proportional to the expected risk premium of the market portfolio.

According to Bodie et all (2013), the CAPM is based on following assumptions:

- The market is perfectly competitive, meaning that there are many investors whose wealth is negligibly small to the total wealth of all investors.
- All investors plan for one identical holding period.
- Investors are exposed to all publicly-traded financial assets and to risk-free borrowing and lending arrangements.
- No taxes and transaction costs.

- All investors are rational, viewing and analyzing securities in the same way and share the same economic view of the world.

If we denote R as return of asset, $R_f$ as risk-free rate, $R_m$ as return of the market portfolio, the CAPM model is usually expressed by the equation:

$$E(R) - R_f = \beta \left( E(R_m) - R_f \right) \quad (1)$$

$$E(R) = R_f + \beta \left( E(R_m) - R_f \right) \quad (2)$$

Note that in this equation, R and $R_m$ are random variables, while $R_f$ is not. The left-hand side of equation (1) represents the risk premium of asset i, while the expression in the parentheses on the right hand side represents the risk premium of the market portfolio. The beta in the equation represents the sensitivity of the asset to a movement in the market. In almost all cases, the beta is positive. A beta greater than 1 implies that the asset is riskier relatively to the overall market, while a beta smaller than 1 implies that the asset is less risky relative to the overall market. For example, a beta of 1.33 means that the asset is 1.33 times more volatile than the overall market. Based on this definition, the beta of the market portfolio is 1.

The traditional theory of finance tells us that a higher risk investment should be compensated by a higher return. Based on this theory, a higher beta – as a measure of higher systematic risk - should generate a higher return in the model. In other words, the CAPM tells us that the beta should be considered as a signal to predict the relative return.

How do we calculate the beta? Because the equation (1) should hold all the times (in theory), we could use historical data to estimate the beta of any assets or portfolio. As we already know the historical return (historical return is not a random variable), we may set up a regression to estimate the beta:

$$R_i - R_f = \alpha + \beta (R_{mi} - R_f) + \varepsilon \ (*) \text{ for all } i = 1,2,\ldots n.$$

If the CAPM equation holds all the time (i.e assets are fairly priced), the alpha (intercept of the regression) should be zero. If alpha is positive (negative), an asset is earning higher (lower) risk premium than expected from CAPM, meaning it is currently undervalued (overvalued). The alpha, therefore, provides an important information for investors to decide whether to buy or sell a financial asset.

## THE FAMA-FRENCH MODEL

The Fama-French model is an extension of the CAPM, first developed by Fama and French (1993). Unlike the CAPM model that uses only one factor (the market portfolio) to explain stock returns, the Fama-French model add two more risk factors in the model. The equation of the model, in the express form, is:

$$R_i - R_f = \alpha + \beta_m (R_{mi} - R_f) + \beta_{smb} \, SMB_i + \beta_{hml} \, HML_i + \varepsilon$$

for all $i = 1,2,\ldots n.$

While R$_m$ has the same meaning as in the CAPM model, SMB stands for "**S**mall **M**inus **B**ig" and HML stands for "**H**igh **M**inus **L**ow." More specifically, stocks are classified by market capitalization (market-cap) and by book-to-market ratio. SMB is the difference in returns between the smallest and largest market-cap stocks, while HML is the difference in return between high and low book-to-market corresponding firms. Based on historical observations,

Fama-French suggested that, small cap stocks usually outperform big cap stocks (size premium), and value stocks (high book-to-market ratio) usually outperform growth stocks (low book-to-market ratio) (value premium). It is not obvious what kind of risks are captured by SMB and HML, but they may "proxy for yet-unknown more fundamental variables" (Bodie et al 2013.) As argued by Fama and French, for example, high book-to-market value may be associated with financial distress, and small stocks are more sensitive to changes in business cycle (Fama and French 1996). Liew and Vassalou (2000) argued that HML or SMB could predict GDP growth, which may capture some aspects of business cycle risk.

In terms of behavioral finance, value premium is a result from market irrationality. One possible explanation is from the recency bias, from which investors are usually excessively optimistic or excessively pessimistic about a stock/firm when looking at the current growth rate. Value stocks (high book-to-market ratio) usually have low current growth rates of earnings/revenues, but investors make a mistake of past low growth for future low growth. Therefore, stock price is likely to beaten down relative to "fundamental value", and when growth surprises on the upside, prices rise leading to higher return. The opposite happens with the "glamour" stock (low book-to-market ratio).

Nonetheless, an important assumption of the Fama-French three factor model, is the risk caused by the size and value/growth are not captured by the market risk factor ($R_m$); otherwise, the explanatory power of the model does not improve much compared to the CAPM.

While many papers concluded that the Fama-French model had a better predictive performance than the CAPM in various settings (Denis 2013, Sehgal and Balakrishnan 2013, etc.), both models are usually tested on the whole data market that contains all stocks of different sizes and values at a time. In the paper, we would like to compare two models' predictive power in a data set of big market-cap stocks to see if the result is consistent with previous studies.

## OVERVIEW OF SOME STATISTICAL METHODS IN ESTIMATING THE CAPM AND THE FAMA-FRENCH MODEL

Mathematically, both the CAPM model and the Fama-French model are all linear regressions (in both parameters and predictors). While the CAPM model is the case of a simple linear regression, the Fama-French model is a multiple linear regression. Therefore, all statistical methods that are used to estimate the CAPM could be applied to estimate the Fama-French model. There are some additional problems, nevertheless, that only arise with the Fama-French model because of more than one predictor.

To avoid lengthy technical description, we only briefly described methods in the case of estimating the CAPM model. When applying these methods to the Fama-French model, statistical assumptions and steps are basically the same; the major difference remains only in the dimension of coefficient vectors and matrices.

### Ordinary Least Square Method (OLS)

The Ordinary Least Square method is the most popular and also the simplest method to estimate coefficients of the linear regression model. Given the equation of the CAPM, the OLS method tries to minimize the sum of squared residuals (RSS), and thus gives the following estimate for beta and alpha:

$$\hat{\beta} = \frac{Cov(R_m - R_f, R_i - R_f)}{Var(R_m - R_f)} = \frac{Cov(R_m, R)}{Var(R_m)}$$

$$\hat{\alpha} = \overline{(R_m - R_f)} - \hat{\beta}\overline{(R - R_f)}$$

(The "hat" implies the value is an estimate, while the "bar" implies the average). Mathematical details for deriving these formulas could be found in the appendix.

If we let $y = \begin{bmatrix} R_1 - R_{f1} \\ R_2 - R_{f2} \\ \dots \\ R_n - R_{fn} \end{bmatrix}$, $X = \begin{bmatrix} 1 & R_{m1} - R_{f1} \\ 1 & R_{m2} - R_{f2} \\ \dots & \dots \\ 1 & R_{mn} - R_{fn} \end{bmatrix}$, $\theta = \begin{bmatrix} \alpha \\ \beta \end{bmatrix}$, and $\varepsilon = \begin{bmatrix} \varepsilon_1 \\ \varepsilon_2 \end{bmatrix}$, then the regression could be expressed in the matrix form as:

$y = X\theta + \varepsilon$   (1)

The OLS estimate is: $\widehat{\theta_{ols}} = (X'X)^{-1}X'y$, in which X' denotes the transpose of X.

According to the Markov's theorem, the above estimate is the BLUE (Best Linear Unbiased Estimator) estimate when the following assumptions could not be rejected:

(1) The mean function is correct, meaning there is a linear relationship between the excess return and the market premium.

(2) Errors are independent and have conditional mean zero.

(3) The conditional variance of errors are constant.

(4) The errors are usually normally distributed, so are returns.

Among above assumptions, the assumption (1) and (3) are the most important ones. If (1) is violated, the estimates are not unbiased, and if (3) is violated, the estimates do not have a minimum variance. Mathematically, the assumption (3) has carried over to the independence and constant conditional variance of responses (in this case, the excess return of asset). Although the normality assumption does not affect the validity of the OLS estimate, it is usually assumed for simplicity and convenience of making inference.

The OLS method is simple and convenience in terms of both ideas and computations. However, the disadvantage of the OLS method is that it does not take into consideration information about the distribution of variables in estimating the coefficients. Therefore, if we do not assume normal distribution (assumption 4), or any other kind of distribution for variables, we could not evaluate how good or bad our estimations are (through the variance of the estimates, for example).

## Maximum Likelihood Method

In the maximum likelihood method, some distributions are imposed on the model, more specifically, on the errors of the model. As usual, the most popular distribution imposed on the errors are the normal distribution. For the normal regression (1), if we assume the errors follow the normal distribution with same mean 0 and same variance $\sigma^2$ (unknown), the likelihood function is:

$$L(\theta, \sigma) = \prod_{i=1}^{n} N(y_t|x_t, \theta, \sigma^2) = (2\pi\sigma^2)^{-n/2} \exp\{\frac{-1}{2\sigma^2}(y - X\theta)'(y - X\theta)\}$$

L is maximized at $\widehat{\theta_{mle}} = (X'X)^{-1}X'y$; therefore, under the normality assumption of errors, the maximum likelihood estimates (MLE) of the coefficients are the same as the OLS estimates. Mathematical details to derive the formula for $\hat{\theta}$ in the Maximum likelihood method could be found in the appendix.

However, in the financial market, it is well-supported that return (therefore the errors) does not follow the normal distribution; instead, empirical studies identify that such distribution is usually heavily-tailed (Fama 1965a). Therefore, some heavily-tailed distributions may be better to model errors and returns on the financial market, including the Student t-distribution, a mixture of normal distribution, or a mixture of normal and Cauchy distribution.

## Robust Bayesian Method

In general, the Bayesian method of estimating the parameters requires some prior belief about their distributions, and after getting the data (hence the likelihood function), we update our belief about the parameters through the posterior distribution. If we denote $\pi(\theta)$ as the prior distribution of $\theta$ (in this situation, $\theta$ is a vector), and $f(x|\theta)$ as the likelihood of the data, then the posterior distribution of $\theta$ is given by:

$$f(\theta|x) = \frac{\pi(\theta)f(x|\theta)}{\int \pi(\theta)f(x|\theta)d\theta}$$

Under the square error loss function, the Bayesian estimator of the parameters is the expected value of the posterior distribution:

$$\hat{\theta} = E(\theta|x)$$

Two challenging aspects of the Bayesian method lies in (1) the robustness and (2) computational efficiency. The result from the Bayesian method may much depend on the prior, but there exists no theoretical rule of how we could choose the best prior. Instead, while theoretical Bayesian approach fixes the prior based on some belief or information in the past, empirical Bayesian approach estimates the prior from the past data. From theoretical approach, the most popular prior employed is the conjugate prior, meaning one that makes the posterior has the same kind of distribution as the chosen prior. While the conjugate prior makes computation and mathematical inference easily, it is usually not robust and sometimes may be irrational (for example, for the return in financial market, it makes no sense to choose a Bernoulli distribution as the prior). For example, if the likelihood of the regression is:

$$L(\theta, \sigma) = \prod_{i=1}^{n} N(y_t | x_t, \theta, \sigma^2) = (2\pi\sigma^2)^{-n/2} \exp\{\frac{-1}{2\sigma^2}(y - X\theta)'(y - X\theta)\}$$

Then it can be shown that the conjugate prior is a normal-inverse-gamma distribution with distribution (Wong and Bian 2000):

$$\pi(\theta, \sigma^2) = \pi(\theta|\sigma)\pi(\sigma^2)$$

where $\pi(\theta|\sigma)$ is the multivariate normal distribution with mean vector $\theta_0$ and precision matrix $V$ (inverse of covariance matrix), and $\pi(\sigma^2)$ is the inverse gamma distribution.

The posterior distribution also follows the normal-inverse-gamma distribution, and the Bayesian estimation of $\theta$, under the squared error loss, is the posterior mean, given by:

$$\widehat{\theta_N} = E(\theta|X, y) = (X'X + V)^{-1}(X'X\widehat{\theta}_{mle} + V\theta_0)$$

The estimator is mathematically nice, because it could be written in a closed form. However, for the CAPM model, $\theta = (\alpha, \beta)$, it is hard to give a specific and meaningful reason for the fact that the beta (representing the volatility of a financial asset in comparison to the market in general) and the alpha (representing how the market values a financial asset in relative to its fair value) have normal-inverse-gamma joint density!

As said above, the financial return may be better described by a heavily-tailed distribution, and so are the coefficients. However, the computation may become extremely ugly, and the posterior distribution may not have a nice form that we could extract the expectation (sometimes, the expectation may not exist as well). To overcome this difficulty, Bian and Dickey (1996) provided the following priors:

$$\pi_p(\theta, \sigma^2) = \pi\ (\theta|g)\pi(\sigma^2)$$

In which $\pi\ (\theta|g)$ is the Cauchy g-prior distribution with mean $\theta_0$ and prior precision g, and $\pi(\sigma^2)$ is an inverse gamma distribution. In this case, with the normal likelihood, the marginal posterior density of $\theta$ is a poly-Cauchy density. Under the squared error loss, the robust Bayesian estimator is the posterior mean and given by:

$$\widehat{\theta_C} = E(\theta|X, y, g) = w\hat{\theta} + (1 - w)\theta_0$$

In which $\hat{\theta} = \widehat{\theta_{ols}} = (X'X)^{-1}X'y$, and $w = (1 + g^{\frac{1}{2}}\|y - X\hat{\theta}\|)^{-1}$

Mathematical derivation of all formulas could be found in Bian and Dickey (1996).

We see that the robust Bayesian estimator is the weighted average of the mean prior and the OLS estimator, and the weight itself is a decreasing function of the prior parameter g and the residual using the OLS method. When an extreme observation occurs to the data, the high value of the residual causes the weight to be smaller, and therefore, the estimator $\widehat{\theta_C}$ is adaptive to such events. In this sense, $\widehat{\theta_C}$ is likely to be more robust.

So there are two remaining questions in choosing the prior:

(1) How to choose the parameter g?

(2) How to choose the prior mean $\theta_0$?

The first question is dealt intensively in the paper of Liang et al (2000), which outlines 5 main approaches to determine the value of g. We will use two following methods in choosing the value of g:

- Benchmark prior: Based on suggestions of Fernandez et al (2001), we could choose $g = max\,(n, p^2)$, in which n is the number of observations and p is the number of regressors (also called the dimension of the model).
- Local empirical Bayes: In this method, g is chosen as $g = max\{F - 1, 0\}$, in which F is the F-statistic for testing the significance of the coefficients of the slope.

The second question is even trickier. In theoretical approach, the value of the prior mean should be fixed, but the robustness could be of serious concern. Hence, the value of $\theta_0$ may be chosen from an empirical Bayesian approach; one practical way is that we could use the estimate of $\theta$ from the previous sample with equal size as the value of prior mean in the updated estimation, as suggested by Martiz and Lwin (1989).

# EMPIRICAL STUDIES

We tested the predictive power of several methods in estimating two models with the data set of big-cap stocks. For both the CAPM and the Fama-French model, the metric we use to compare its predictive power is the mean squared error on specific subsets of data using Leave-one-out cross validation and out-of-sample prediction. For the CAPM model, we also test if beta coefficients, estimated by any method, could be a good signal of returns ranking of stocks; in other words, if we see stock A has a higher beta than stock B, does A usually have higher return than B?

## Data

The data we used to test both models were from the historical monthly return of stocks in S&P 500 index from July 2010 to December 2014 at the time of January 15, 2015 (number of period n =42). For the robust Bayesian method, we also needed the return of stocks from January 2007 to June 2010 to get the prior mean (the size of previous sample should be equal to the main data in the model). All return data of stocks and market portfolio were available from the Center for Research in Security Prices (CRSP) at the University of Chicago. To track the stock and make sure one stock corresponds only to one business in the period, we included its Ticker, permanent identifiers (PERMANO), and CUSIP number in the downloaded file. However, because all these factors could be changed and be reused over time (for example, one ticker at two different points of time may belong to two different stocks), we had to designate an additional factor (called it IDENTTY) based on all three above factors to make sure each stock has a unique IDENTITY. The value of IDENTITY is similar to the ticker, but not necessarily exactly the same. Each stock, therefore, corresponded to one IDENTITY, had historical returns of 84 periods; we removed any stock that had fewer than 84 returns. After clearing the data, we had 468 stocks, ordered alphabetically from A to ZMH; then, we cut the data into two sub-data: The Late data containing historical returns from July 2010 to December 2014 was the primary data for model comparison, while the Early data contained historical returns from January 2007 to July 2010 were used to get the prior mean in the Bayesian method only.

The risk-free rate is taken from the historical data of the 3 month T-bill rate, available from the Economic Research of Federal Reserve Bank in St. Louis. The data for Fama-French factors corresponding to the above periods are available from the data library of Kenneth R. French at the Kenneth R. French's data library.

Originally, all original return data were discrete. We also converted them into continuous, and then ran regressions on both types of data to see if there had been a significant difference in result. All regressions were implemented on R.

## Computational Procedure and Result

### Estimating the CAPM model by the OLS method

For computational efficiency, instead of running CAPM on every single stock, we ran a complex regression that included the Risk premium $(R - R_f)$ as response, and both Market premium $(R_m - R_f)$ and IDENTITY and their interactions as predictors. Noted that IDENTITY is a categorical variable, and we used it to keep track of results for single stocks.

Based on the OLS method, we ran the CAPM model twice, one based on the discrete return and one based on the continuous return. We got the alpha and beta of the first stock (A) directly, while the coefficients of interactions terms revealed the difference of alphas and betas of all remaining stock from the alpha and beta of the first stock respectively. The following table summarized betas of stocks getting from the OLS, both discrete and continuous cases.

|  | Betas of stocks based on continuous returns | Betas of stocks based on discrete returns |
| --- | --- | --- |
| Minimum | -0.2240 | -0.2350 |
| $1^{st}$. quarter | 0.6838 | 0.6943 |
| Median | 1.0378 | 1.0413 |
| Mean | 1.0879 | 1.0880 |
| $3^{rd}$. quarter | 1.4775 | 1.4922 |
| Max | 2.8250 | 2.7500 |

Table 1. Descriptive statistics of betas based on the OLS method.

The histogram of betas estimated from the OLS showed that most betas (estimated from either discrete or continuous) concentrated heavily around 1.

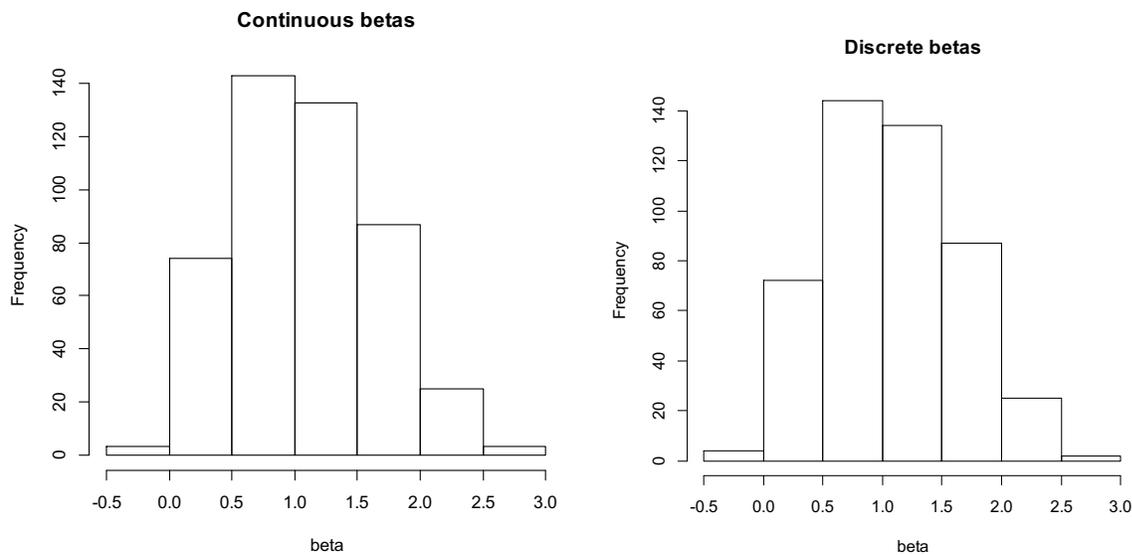

Figure 2. Histograms of betas based on the OLS method.

Next, we ranked stocks based on betas. Top 5% stocks and bottom 5% stocks based on betas for both cases were shown in the table below:

| Top 5% beta (Continuous) | NBR, STX, MWW, ETFC, MS, FFIV, FCX, JNPR, DNR, RF, JOY, ATI, BTU, THC, LNC, FOSL, VLO, GNW, OI (largest to smallest) |
|---|---|
| Top 5% beta (Discrete) | NBR, STX, MWW, FFIV, MS, ETFC, JNPR, FCX, JOY,THC, DNR, VLO, RF,TSO, GT, CBG, HAR, ATI, NOV (largest to smallest) |
| Bottom 5% beta (continuous) | PEG, AEE, XEL, VRTX, ACT, BMY, DLTR, WEC, PRGO, HSY, D, CMS, GIS, DUK, ETR, FE, KMB, SO, ED, NEM, MNST (largest to smallest) |
| Bottom 5% beta (discrete) | PEG, XEL, AEE, BMY, ACT, PRGO, WEC, DLTR, HSY, D, CMS, GIS, DUK, FE, ETR, KMB, SO, VRTX, ED, NEM, MNST (largest to smallest) |

*Table 2.Top 5% and Bottom 5% Stocks according to betas based on the OLS method.*
.

There is no significant difference in the top and the bottom of the ranking between using continuous return and using discrete returns. Almost all stocks in the top (bottom) 5% of betas discrete also lied in the top (bottom) 5% of betas continuous, and vice versa.

Next, we examined if the ranking of historical returns mimicked the ranking of betas to examine if a higher beta was associated with a higher historical return. Because most betas concentrated heavily in one small interval, so the actual difference among betas might not be significant to conclude any relationship or association. Therefore, we had better to examine the relationship in extreme cases (top and bottom 5%) and in general. First, we computed the geometric average discrete return and geometric average continuous return of each stock over the chosen period.

|            | Geometric average discrete returns | Geometric average continuous returns |
|---|---|---|
| Minimum    | -0.02113 | -0.03907 |
| 1st. quarter | 0.01060 | 0.00820 |
| Median     | 0.01604 | 0.01359 |
| Mean       | 0.01604 | 0.01286 |
| 3rd. quarter | 0.02166 | 0.01901 |
| Max        | 0.06163 | 0.05257 |

*Table 3. Descriptive statistics of historical geometric average returns.*

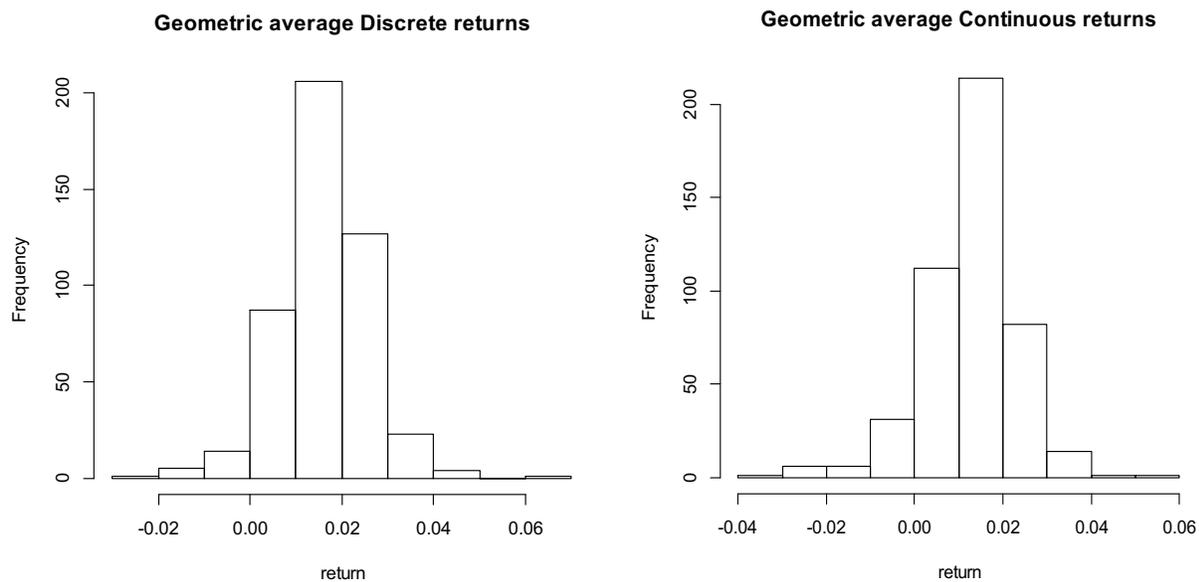

*Figure 3. Histogram of historical geometric average returns.*

The table and the histograms showed that most stocks in the data set had geometric average return positive, but less than 3%. The difference between maximum and minimum geometric return in both cases, however, were quite large, more than 8%. In both cases, the geometric average returns did not seem to follow a normal distribution.

The below table shows top and bottom 5% stocks based on geometric average return:

| | |
|---|---|
| Top 5% return (Continuous) | REGN, PCLN, BIIB, TSCO, ALXN, UA, CBS, ADS, COG, MA, GILD, ACT, STZX, STZ, TSO, MCO, WYN, LTD, VFC, AMZN, PPG |
| Top 5% return (Discrete) | REGN, PCLN, BIIB, UA, TSCO, ALXN, TSO, CBS, COG, STX, STZX, STZ, ADS, GILD, MA, MCO, ACT, WYN, CMG, LTD, EXPE |
| Bottom 5% return (continuous) | BAC, FTR, EXC, BRCM, SPLS, HCBK, JNPR, COHU, ATI, NBR, HSP, AVP, X, HPQ, NFX, MPET, EBIX, NEM, BTU, MWW, FSLR |
| Bottom 5% return discrete | SWN, JNPR, NBR, EBIX, BRCM, COHU, SPLS, FTR, HCBK EXC, ATI, X, AVP, HSP, HPQ, MWW, MPET, BTU, NFX, FSLR, NEM |

*Table 4.Top 5% and Bottom 5% Stocks based on historical geometric average return.*

Comparing two tables, we saw that no stock in the top 5% betas stayed in the top 5% geometric average return (in both discrete and continuous cases), and no stock in the bottom 5% betas stayed in the bottom 5% geometric average return. We concluded that the beta is not a good indicator for stocks' return in extreme cases.

To see if beta still works overall, we examined a correlation between ranking of beta and ranking of geometric average return. For each stock, corresponding to each IDENTITY, we assigned its beta rank (X), and its geometric average return's rank (Y). In the discrete case, the correlation between X and Y is -0.06, while in the continuous case, the correlation between X and Y is -0.20.

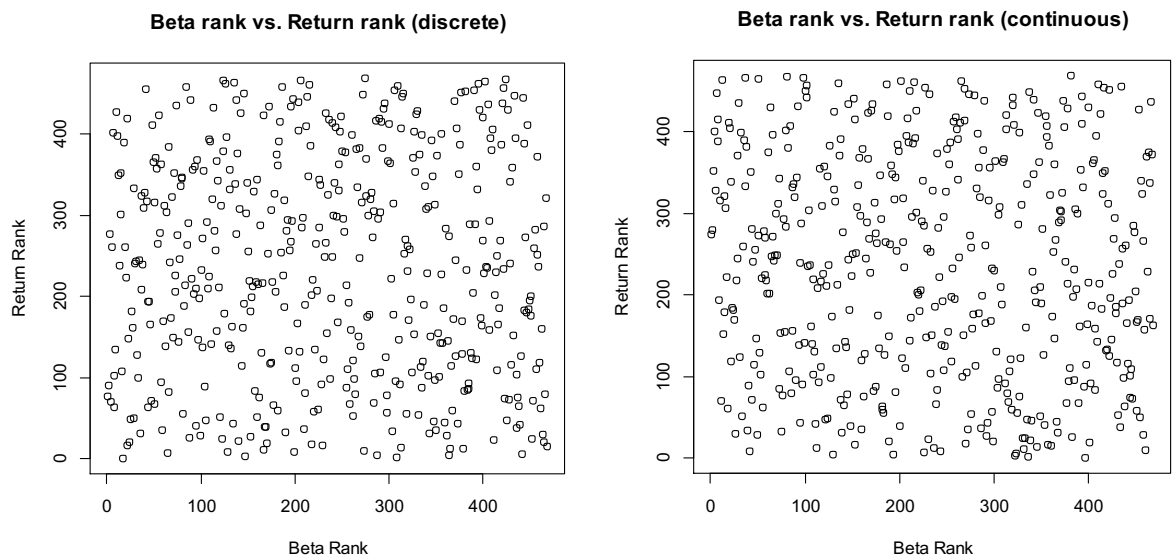

*Figure 4. Ranking of stocks based on beta estimated by the OLS method and Ranking of stocks based on historical geometric average return.*

The above graphs show no pattern of the beta rank and the return rank, so we concluded that overall, there is no relationship between the beta and the past historical return in the data. In other words, the CAPM model did not work well in the data set.

Normality test and Estimating the CAPM model by the Maximum Likelihood method

We knew that the estimator of coefficients based on the OLS method was the same as the estimator of coefficients based on the maximum likelihood method (MLE) when the return was normally distributed. In this section, we tested the normality of historical returns to see if the MLE and the OLS gave the same estimates.

We tested the normality based on the Shapiro-Wilk test for both the historical discrete and continuous cases. For the discrete cases, out of 468 stocks, 414 stocks generated p-values greater or equal to 0.05; for the continuous case, out of 468 stocks, 403 stocks generated p-values greater or equal to 0.05. So at this significance level, the Shapiro-Wilk test retained normality for the majority of stocks' historical return.

Hence, for the majority of stocks, we may consider the beta estimated by the OLS method also the maximum likelihood estimator of the beta.

Estimating the CAPM model by the robust Bayesian method

Based on the robust Bayesian method, the estimator of coefficients are the weighted average of the OLS coefficients and the chosen priors, and the weights depend on the residuals of the OLS method. We implemented the robust Bayesian method in the data set. According to the empirical Bayesian approach, the prior mean $\beta_0$ could be chosen as the estimate of beta from the OLS method from the data of historical return from January 2007 to June 2010. The parameter $g$ could be chosen through the benchmark prior or local empirical Bayes approach; we implemented both options. Note that, in the benchmark prior approach, g is equal to 42 for all stocks (g=max (n,p)), but in the local empirical approach, the value of g varied from stock to stock. The computational results were showed below.

First, we looked at some descriptive statistics of the betas estimated in both cases of g.

Case 1: g=42 (benchmark prior)

|  | Betas of stocks based on continuous returns | Betas of stocks based on discrete returns |
|---|---|---|
| Minimum | 0.6742 | 0.184102 |
| 1st. quarter | 1.08966 | 0.760846 |
| Median | 1.236154 | 1.075593 |
| Mean | 1.2312 | 1.1202 |
| 3rd. quarter | 1.384744 | 1.38046 |
| Max | 2.063182 | 4.21851 |

Table 5. Descriptive Statistics of estimated betas based on robust Bayesian method with benchmark priors

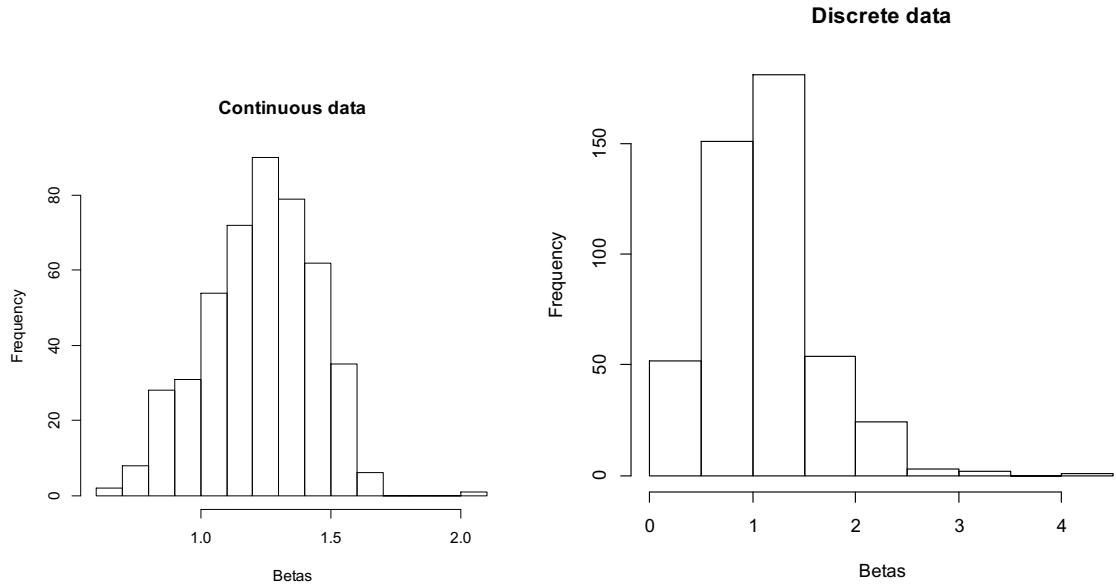

Figure 5. Histograms of betas estimated by the robust Bayesian method with benchmark priors.

Local Bayes approach

|  | Betas of stocks based on continuous returns | Betas of stocks based on discrete returns |
|---|---|---|
| Minimum | 0.045613 | -0.02313 |
| 1$^{st}$. quarter | 1.006615 | 0.75136 |
| Median | 1.217266 | 1.042912 |
| Mean | 1.1491 | 1.0952 |
| 3$^{rd}$. quarter | 1.388156 | 1.373481 |
| Max | 2.07897 | 4.059774 |

Table 6. Descriptive Statistics of betas based on the robust Bayesian method with local empirical Bayes prior.

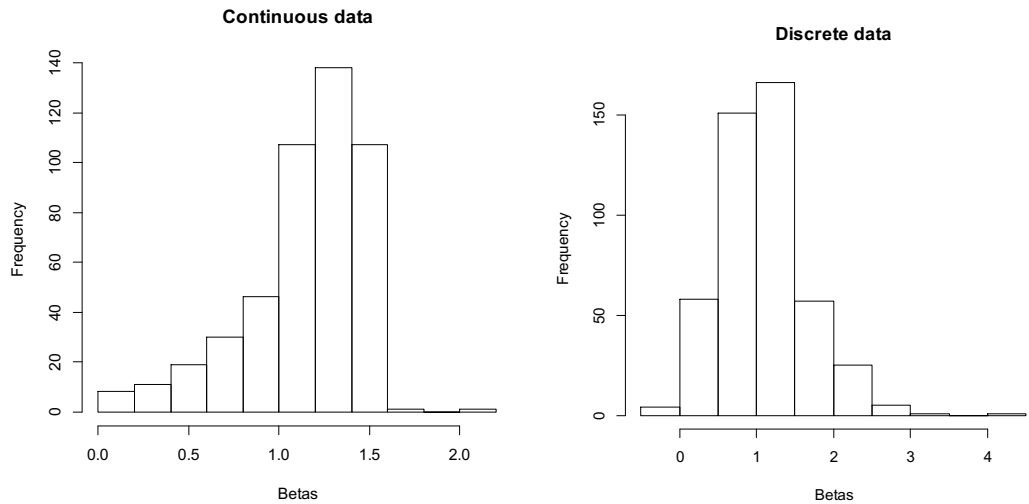

*Figure 6. Histograms of betas based on the robust Bayesian method with local empirical Bayes prior.*

We might see that the range of betas using robust Bayesian method in discrete cases (based on both local Bayes and benchmark prior approaches) were large, where the maximum beta was around 4. This was caused by the extremely high prior mean (approximately 5.115) that was not neutralized by any extremity in the value of F-statistic and/or the residual.

Could these estimated betas have some relationship with historical returns? Similar to the OLS method, we established the rankings of stocks according to betas and according to returns. The tables below demonstrated top 5% and bottom 5% according to betas estimated by the robust Bayesian method:

Robust Bayesian method with benchmark prior:

| Top 5% beta (Continuous) | AA, NBR, MS, ETFC, LNC, IVZ, DNR, OI, FCX, HIG, RF, STX, ATI, MET, C, JNPR, NOV, JOY, LUK, CBG |
|---|---|
| Top 5% beta (Discrete) | PIR, GNW, AIG, HIG, WYN, LNC, PFG, GCI, GT, TXT, CBG, THC, F, WYNN, X, C MAC, PRU, HST, XL, BAC, DOW |
| Bottom 5% beta (Continuous) | SO, ED, DUK, D, WEC, KMB, GIS, XEL, MCD, CMS, HSY, SCG, GAS, DTE, CL, PEP, PCG, POM, ETR, KO |

| Bottom 5% beta (Discrete) | ED, SO, GIS, FDO, SRCL, WEC, DUK, WMT, KMB, HSY, PCG, PPL, FE, XEL, CPB, D, ABT, MCD, BCR, DLTR, PEG |

*Table 7. Top 5% and Bottom 5% Stocks according to betas based on the robust Bayesian method with benchmark prior.*

Robust Bayesian method with Local empirical Bayes:

| Top 5% beta (Continuous) | AA, NBR, STX, MWW, ETFC, FCX, MS, DNR, FSLR, JNPR, RF, OI, THC, LNC, JOY, BTU, FFIV, ATI, GNW, NOV, TSO |
|---|---|
| Top 5% beta (Discrete) | PIR, GNW, AIG, HIG, WYN, PFG, LNC, MAC, CBG, GT, TXT, GCI, XL, THC, C, PRU, HST, WYNN, F, X, STX |
| Bottom 5% beta (Continuous) | ED, SO, NEM, KMB, DUK, FE, MNST, ETR, GIS, CMS, D, HSY, WEC, PRGO, DLTR, ACT, BMY, VRTX, XEL, AEE, POM |
| Bottom 5% beta (Discrete) | MNST, NEM, VRTX, ED, SO, KMB, GIS, FE, DUK, ETR, HSY, DLTR, WEC, CMS, ACT, D, PRGO, BMY, XEL, PPL, PCG |

*Table 8. Top 5% and Bottom 5% of stocks according to betas based on the robust Bayesian method with local empirical Bayes prior.*

The above tables demonstrated that although prior parameters could be chosen differently, the list of stocks that made top 5% and bottom 5% in the same kind of data (discrete or continuous) were almost the same. However, there were a significant difference in the ranks of stocks in discrete and continuous cases; for instance, there were only few stocks in the list of top 5% beta in continuous case that were also in the list of top 5% beta in discrete case.

Nevertheless, when comparing with the ranks of betas based on the OLS method (Table ), there was a major difference in the discrete and continuous case. In the continuous case, all rank (at the extremes – top and bottom 5%) were similar. In the discrete case, there was a significant difference between the ranks from local empirical Bayes with the two other ranks. For example, most stocks in the top 5% beta based on OLS method were also in the top 5% beta based on the benchmark prior (the ranking of other stocks was also close to the top 5%), while they were pulled down a lot in the rankings based on local empirical Bayes. Similarly, most stocks in the top 5% beta based on local empirical Bayes were not in the top of neither the ranking based on

OLS nor the ranking based on the benchmark prior. Therefore, we might see that in the discrete case, the estimated coefficients in the local empirical Bayes were much adjusted for the impact of fitness of the CAPM model (through the residual) and the parameters of priors.

Comparing the ranks of stocks according to betas based on robust Bayesian methods with the ranks of stocks according to geometric average return in the Table 4, we still saw that in both discrete and continuous cases, no matter how the prior was chosen, no stock in the top (bottom) 5% beta stayed in the top (bottom) 5% geometric average return. The following table summarized the correlations among the ranks of stocks according to beta and according to geometric average return from the CAPM model using different methods of estimation.

Continuous case:

|  | *Rank beta OLS* | *Rank beta benchmark prior* | *Rank beta local empirical Bayes* | *Rank geometric average return* |
|---|---|---|---|---|
| Rank beta OLS | 1 | | | |
| Rank beta benchmark prior | 0.976305407 | 1 | | |
| Rank beta local empirical Bayes | 0.997197368 | 0.983665689 | 1 | |
| Rank geometric average return | -0.204873538 | -0.197641344 | -0.19928 | 1 |

Discrete case:

|  | *Rank beta OLS* | *Rank beta benchmark prior* | *Rank beta local empirical Bayes* | *Rank geometric average return* |
|---|---|---|---|---|
| Rank beta OLS | 1 | | | |
| Rank beta benchmark prior | 0.976305407 | 1 | | |
| Rank beta local empirical Bayes | 0.852170672 | 0.864443226 | 1 | |
| Rank geometric average return | -0.065652819 | -0.042140054 | 0.032363186 | 1 |

*Table 9. Correlation among betas estimated by different statistical methods.*

The above empirical analysis showed that although three methods generated three different rankings, they were highly positively correlated; therefore, the relationship of beta and return in all cases were essentially the same. However, all of the beta rankings had very low correlation with the ranking of geometric average return in general. For the data, although the

CAPM model could be estimated by different statistical methods, we concluded that the beta was not a good indicator of return in extreme cases or in general.

## Comparing the predictive power of the Fama-French model and the CAPM model

Recall that the regression equation of the Fama-French model is:

$$R - R_f = \alpha + \beta_m(R_m - R_f) + \beta_{smb}\, SMB + \beta_{hml}\, HML$$

Because the Fama-French model is an extension of the CAPM model in terms of additional risk factors, the statistical methods used above to estimate the CAPM model could be applied to estimate the Fama-French model. The only major difference is, while the CAPM model could give us some potential indicators (beta) to predict the return, it is hard to extract any indicator from the Fama-French model separately. Instead of having one slope that could be compared directly, the Fama-French model generates a vector of three slopes, which is challenged to develop a consistent set of criteria for comparison. Therefore, we compared two models using their predictive performance.

Statistically, the Fama-French model adds two more predictors (HML and SMB) into the CAPM model; therefore, the Fama-French model should explain the variability of the response (the excess return) better than the CAPM model (coefficient of determination $R^2$ of the Fama-French model is higher). However, more predictor does not guarantee a higher predictive power in general; by using predictive power, we had more objective criteria to evaluate how good a model is. We did not have to make any assumption on the distribution of the data to evaluate the model; however, it would be more challenging to explain why a model is better another model theoretically, and to examine if one model is better in general or just in a particular data set.

To compare the predictive power of both CAPM and Fama-French model, we used two following methods. First, we ran the model in the Early Data to generate coefficients, and then used the model to make prediction on the Late Data (out-of-sample prediction). We tracked the error of each prediction and compared the mean squared errors of different models.

The second method we used was Leave-one-out cross validation, a widely-used computationally-intense method for model comparison. We divided the LateData into 468 sub-data, each of which corresponded to a unique IDENTITY, then ran the model on each of the sub-data. Then, we ran leave-one-out cross validation for the CAPM and for the Fama-French model in each sub-data, whose general procedure in a data set containing K data points was described below:

- For each point $i = 1, 2, \ldots, K$:
    - Remove point $i^{th}$ of the data. The set of remaining points is called a training set.
    - Run the model (CAPM or Fama-French) on the created training set
    - Use the model to predict and calculate the prediction error for the $i^{th}$ point.
- After having prediction errors on all K points, calculate the mean of all the squared prediction errors on the subdata.

Because the above procedure was run for all 468 sub-data, for each model (CAPM and Fama-French), we had a vector of length 468 that contained all mean squared prediction errors from each sub-data. We compared some descriptive statistics of these vectors.

Out-of-Sample Prediction Performance

The following table summarized the out-of-sample prediction errors on 468 sub-data (each sub-data corresponds to one IDENTITY) of the CAPM model and the Fama-French model (both estimated by the OLS method).

Discrete case:

| Error Information | The CAPM model | The Fama-French model |
| --- | --- | --- |
| Min of mean squared error | 0.0006984 | 0.000649 |
| Mean of mean squared error | 0.0049150 | 0.005098 |
| Max of mean squared error | 0.0674200 | 0.050150 |
| Range of mean squared error | 0.0667216 | 0.049501 |

Continuous case:

| Error Information | The CAPM model | The Fama-French model |
| --- | --- | --- |
| Min of mean squared error | 0.0006621 | 0.000613 |
| Mean of mean squared error | 0.0044960 | 0.004807 |
| Max of mean squared error | 0.0466400 | 0.045100 |
| Range of mean squared error | 0.0459779 | 0.044487 |

*Table 10.Descriptive statistics of mean squared errors of out-of-sample prediction*

In both discrete and continuous cases, the CAPM model had a lower mean of prediction error, but the Fama-French model has a narrower range of error. The inconsistent comparison told us that no model was better than the other in predicting out-of-sample excess return with this data.

Leave-one-out cross validation

The following table summarized prediction errors using Leave-one-out cross validation on each of the 468 sub-data (each sub-data corresponds to one IDENTITY) of the CAPM model and the Fama-French model (both estimated by the OLS method).

Discrete case:

| Error Information | The CAPM model | The Fama-French model |
|---|---|---|
| Min of mean squared error | 0.0003235 | 0.0001894 |
| Mean of mean squared error | 0.0043720 | 0.0045170 |
| Max of mean squared error | 0.0406600 | 0.0447000 |
| Range of mean squared error | 0.0403365 | 0.0445106 |

Continuous case:

| Error Information | The CAPM model | The Fama-French model |
|---|---|---|
| Min of mean squared error | 0.000434 | 0.000457 |
| Mean of mean squared error | 0.004191 | 0.004318 |
| Max of mean squared error | 0.057500 | 0.057400 |
| Range of mean squared error | 0.057065 | 0.056957 |

*Table 11.Descriptive statistics of mean squared errors of Leave-one-out cross validation.*

Similar to the result from the out-of-sample prediction performance, the result from Leave-one-out cross validation showed that the range of prediction error of the CAPM model was wider, while the mean of the error of the Fama-French model was higher. The inconsistency in the comparison demonstrated that there was no clear evidence to support that the predictive power of one model was better than the other.

All empirical analyses demonstrated that, the Fama-French model has not worked better than the CAPM model in predicting return of large market-cap stocks. The inconsistency of this result compared with the results from previous studies may reveal some weakness of the Fama-French model.

## WHY DOES THE CAPM FAIL?

In the above section, we have discussed quantitatively that the theoretical CAPM failed to predict the relative return of stocks in the data; the beta in the CAPM model is not a good indicator for the high or low future stock return. In fact, the failure of the CAPM not only applied to this specific data, but also was tested in many different data in different settings (though in most settings, the CAPM was only estimated by the OLS method). In this section, we reviewed some of the pitfalls that may prevent CAPM from holding true in reality.

### Leverage constraints

According to Frazzini and Pedersen (2013), a basic condition of the CAPM model is that all investors can use leverage to suit their preferences of risk in investment. Among them, some investors with low-risk aversion, who want high expected return, so theoretically from the CAPM model, these people just need to lever up the Market Portfolio, i.e, they can borrow at the risk-free rate $r_f$ and invest their own wealth and borrowings in the Market Portfolio. However, in the real world, many investors, including individuals, pension funds, and mutual funds, cannot borrow; in other words, they may be constrained in the leverage that they make take. Instead, they deviate from the Market Portfolio by holding less low beta stocks and hold more high beta stocks; therefore, the "artificial" lack of demand for low beta stocks depress their prices, while the "artificial" excess demand for high beta stocks inflate their prices. The change in prices due to the constraints in leverage may provide a reason for why the CAPM failed in reality.

### Some behavioral explanations

According to Kahneman (2002), our brains are divided into two systems when coming to decision making: While system 1 responds by making quick associations, and is fast, intuitive, automatic, unconscious, and effortless, System 2 responds more consciously, but is slow, controlled, deliberate, and suspicious and statistical. System 1 is gullible and prone to mistakes, whereas system 2, though less likely to make mistakes, is costly to use. Kahneman argues that "financial decisions are made in situations of high complexity and high uncertainty that preclude reliance of fixed rules and compel the decision-maker to rely on intuition." In other words, most investors use System 1 in making financial decision, giving rise to pricing error.

One behavioral explanation for the failure of the CAPM model is on the basis of base-rate neglect (or base-rate fallacy), a type error usually made by investors using System 1. One hypothesis is that, when trying to think of "great investments" that provide a high expected return, investors focus on buying stocks whose companies invested in new technologies or products. The road to riches, according to these investors, would be paved with these speculative investments; nevertheless, they ignore the (very large) base rate at which new, small, speculative investments in technologies fail. Therefore, they typically end up overpaying for volatile stocks.

Another (less convincing) explanation comes from "lottery ticket." Many investors consider some types of volatile stocks like buying a lottery ticket in the hope of getting big. However, there is a small chance of doubling or tripling the capital, compared to a much larger chance of a small loss. Technically, a large number of people are overpaying for the small chance of hitting big, which affects price and return of risky financial assets.

## Review of the assumptions of the CAPM model

Finally, we have reviewed assumptions of the CAPM model, which are arguably totally unrealistic. In addition to the leverage constraints noted above, there were many transactions costs (like bid-ask spread, commissions) and taxes associated with investment. In terms of taxes, different investors may face different tax rates, so tax advantage may also be a consideration in deciding the portfolio or stocks investors would like to hold (Brennan 1970). Secondly, while in the CAPM model, expected return comes as a single number for each stock that fulfills market efficiency, regardless of investment's goals. In reality, each investor may have a different expectation, and there is no set of "true expectations" revealed to any market participant that could be used to define an efficient portfolio (Rosenberg 2013). Investors also have diverse goals; some of them hold some financial assets involuntarily. For example, some investors have to hold a company's share because of retention of voting rights and/or incentive compensation for management. Even in some cases, outstanding shares may be closely held in a small group of investors, not widely traded in the public for some period.

In conclusion, the failure of the CAPM in reality demonstrates the inefficiency of the market. Therefore, active investment management is widely practical; "smart" investment strategies still create abnormal returns for investors.

## LIMITATION OF THE FAMA-FRENCH FACTOR MODEL

The above empirical studies demonstrated that in predicting the return of large market-cap stocks, the Fama-French model was not better than the CAPM in the predictive power. Therefore, as consistent with findings of Suh (2009) for individual stocks, we found that the market index (the Market portfolio) was still the most important factor in explaining the return of individual stocks in the data. The effect of additional risk factors including SMB and HML is not significant; instead, it could be more significant on mid and micro-cap stocks or on the whole market overall. One potential explanation, as argued by Vassalou and Xing (2004), is that the size effect (SMB) and the book-to-market effect (HML) exist only within one and two quintiles with the highest default risk or financial distress (the risk that a company could not make payment on its debt obligation). Therefore, for the big stocks in the S&P 500 index, of which most investors do not expect a high default risk, SMB and HML is of little influence.

Aforementioned, the nature of the size effect and book-to-market effect on stock return is not clear. It is controversial that they are responsible for the effect, or just a proxy for unknown factors correlated with them. Most reasons that favor these effects come from investors' behaviors, including overreaction and biasedness, which may be changed and sensitive to individual investors and time. Different empirical studies may generate different results about testing these effects (for example, Kadiyala's study (1995) rejected the hypothesis that book-to-market effect derived from over-reaction, while Nagel (2001) retained this hypothesis) Hence, size and book-to-market effect may be not robust to the data.

# CONCLUSION

Through an empirical study with the historical return of stocks in the S&P 500, we have demonstrated that although it could be estimated under different statistical methods like Ordinary Least Square (OLS), robust Bayesian method (with two different choice for priors), the Capital Asset Pricing Model failed to explain the historical return both absolutely and relatively. In fact, no matter what statistical methods were used to estimate, the CAPM beta, serving as a representation of systematic risk, was not a good signal of return, and the rankings of stocks according to betas were highly correlated with each other but had low correlation with the ranking of stocks according to geometric average returns. The result led us to have more evidence to support the incorrectness of the nature of the CAPM; its unrealistic assumptions, behavior of investors, and leverage constraints are three prominent factors keeping CAPM far from being held in the reality.

The thesis also compared the predictive power of the CAPM and the Fama-French model in predicting the return of big stocks in the market. Using out-of-sample and Leave-one-out cross validation, we concluded that the Fama-French model, although including more risk-factors than the CAPM, had no better predictability than the CAPM in this case. The result demonstrated that additional factors in the Fama-French model like SML and HMB, which reflected the size effect and book-to-market effect respectively, had little effect on predicting return of the big stocks. Therefore, the thesis provided another evidence to support the fact that these effects might only exist and make a difference in micro- and mid-cap stocks, which usually have a high default risk in the overall market.

# APPENDIX

## Mathematical derivation to estimate the CAPM model by the maximum likelihood method

For the normal linear regression with response $y$ and predictor $X$ (in matrix form), if we assume the errors follow the normal distribution with same mean 0 and same variance $\sigma^2$ (unknown) the likelihood function is:

$$L(\theta, \sigma) = \prod_{i=1}^{n} N(y_t|x_t, \theta, \sigma^2) = (2\pi\sigma^2)^{-n/2} \exp\{\frac{-1}{2\sigma^2}(y - X\theta)'(y - X\theta)\}$$

To maximize L, we first compute its logarithm, then take the partial derivatives and set them equals zero.

$$\log L = -\frac{n}{2}\ln(2\pi) - \frac{n}{2}\ln(\sigma^2) - \frac{1}{2\sigma^2}\ln(y - X\theta)'(y - X\theta)$$

$$\frac{d\log L}{d\theta} = \frac{1}{\sigma^2}(y - X\theta)'X = 0$$

$$\rightarrow (y - X\theta)'X = 0 \rightarrow X'(y - X\theta) = 0$$

$$\rightarrow X'y = X'X\theta \rightarrow \widehat{\theta_{mle}} = (X'X)^{-1}X'y$$

## Mathematical derivation to estimate the CAPM model by the Ordinary Least Square (OLS) method

The equation of the regression is:

$$R_i - R_f = \alpha + \beta(R_{mi} - R_f) + \varepsilon \quad (*)$$

To simplify the notation, let $Y_i = R_i - R_f$ and $X_i = R_{mi} - R_f$ for all i= 1,2,…n

The OLS method tries to find $\alpha$ and $\beta$ that minimize the residual sum of squared error

$$RSS(\alpha, \beta) = \sum_{i=1}^{n}(Y_i - \alpha - \beta X_i)^2$$

RSS is a function of both $\alpha$ and $\beta$, so in order to minimize it, we take the derivative with respect to both $\alpha$ and $\beta$:

$$\frac{dRSS}{d\alpha} = -2\sum_{i=1}^{n}(Y_i - \alpha - \beta X_i) = 0 \quad (1)$$

$$\frac{dRSS}{d\beta} = -2\sum_{i=1}^{n}(Y_i - \alpha - \beta X_i)X_i = 0 \quad (2)$$

$$(1) \rightarrow n\alpha = \sum_{i=1}^{n}(Y_i - \beta X_i) \rightarrow \alpha = \overline{Y} - \beta \overline{X} \quad (*), \quad \overline{Y} = \frac{1}{n}\sum_{i=1}^{n}Y_i, \quad \overline{X} = \frac{1}{n}\sum_{i=1}^{n}X_i$$

Plug (*) into (2), we have:

$$\sum_{i=1}^{n}(Y_i - \overline{Y} - \beta(X_i - \overline{X}))X_i = 0 \rightarrow \beta = \frac{\sum_{i=1}^{n}(Y_i - \overline{Y})X_i}{\sum_{i=1}^{n}(X_i - \overline{X})X_i} = \frac{Cov(X,Y)}{Var(X)}$$

Returning the original variables, we have:

$$\hat{\beta} = \frac{Cov(R_m - R_f, R - R_f)}{Var(R_m - R_f)} = \frac{Cov(R_m, R)}{Var(R_m)}$$

$$\hat{\alpha} = \overline{(R_m - R_f)} - \hat{\beta}\overline{(R - R_f)}$$